\documentclass{aa}
\usepackage{txfonts}
\usepackage{xcolor}
\usepackage{natbib}
\bibpunct{(}{)}{;}{a}{}{,}
\usepackage{graphicx}
\usepackage{tabularx}
\usepackage{placeins}
\usepackage[]{rotating}
\usepackage{hyperref}
\hypersetup{
    colorlinks=true,
    linkcolor=blue,
    citecolor=blue,
    }
\usepackage{longtable}
\usepackage{booktabs}
\usepackage{lscape}
\usepackage{ulem}

\newcommand{\lL}{\ifmmode \log \frac{L}{L_{\sun}} \else $\log \frac{L}{L_{\sun}}$\fi}

\newcommand{\xshooter}{{\sc XSHOOTER}}

\newcommand{\pollux}{{\sc POLLUX}}
\newcommand{\vlt}{{\sc VLT}}
\newcommand{\hst}{{\sc HST}}

\newcommand{\mdot}{$\dot{\rm M}$}

\newcommand{\teff}{T$_{\rm eff}$}

\newcommand{\logg}{$\log g$}
\newcommand{\logL}{$\log L/L_\odot$}

\newcommand{\vinf}{$\rm v_{\infty}$}

\newcommand{\kms}{km~s$^{-1}$}

\newcommand{\zsun}{${\rm Z}_{\odot}$}

\newcommand{\cmfgen}{{\sc CMFGEN}}

\newcommand{\powr}{{\sc PoWR}}
\newcommand{\ullyses}{{\sc ULLYSES}}
\newcommand{\jwst}{{\sc JWST}}
\newcommand{\xshootu}{{\sc XShootU}}
\newcommand{\MONEMC}{{\sc 1001MC}}
\newcommand{\bloem}{{\sc BLO{\rm e}M}}

\begin{document}

 \title{CMFGEN grids of atmosphere models for massive stars}
\subtitle{OB-type stars at the Magellanic Clouds}%
\author{
      {W. Marcolino\inst{\ref{ONR}}
      }
            \and 
      {J.-C. Bouret\inst{\ref{LAM}}}
      \and
      {F. Martins\inst{\ref{LUPM}}}      
       \and
      {D.J Hillier\inst{\ref{PITT}}}
}
\institute{
    {Universidade Federal do Rio de Janeiro, Observat\'orio do Valongo. Ladeira Pedro Ant\^onio, 43, CEP 20080-090,
           Rio de Janeiro, Brazil\label{ONR}}
   \and
      {Aix Marseille Univ, CNRS, CNES, LAM, Marseille, France\label{LAM}}
   \and   
   {LUPM, Universit\'e de Montpellier, CNRS, Place Eug\`ene Bataillon, F-34095 Montpellier, France \label{LUPM}} 
   \and
   {Department of Physics and Astronomy \& Pittsburgh Particle physics, Astrophysics, and Cosmology Center (PITT PACC), University of Pittsburgh,  Pittsburgh, PA 15260, USA\label{PITT}} 
}

\offprints{Wagner Marcolino\\ \email{wagner@ov.ufrj.br}}

\date{Received / Accepted }

\abstract
{Large spectroscopic surveys of individual massive stars, such as \ullyses\ and \xshootu, provide observational data for hundreds of massive stars. Their analysis requires large numbers of synthetic spectra so that stellar parameters can be determined. In addition, libraries of massive stars' spectra are needed to produce population synthesis models able to reproduce the observed spectra of unresolved young stellar populations, such as those revealed by the James Webb Space Telescope (\jwst) in the early Universe.}
{Our main goal is to provide an extensive library of synthetic spectra and spectral energy distributions of OB stars at metallicities of the Magellanic Clouds. This library will offer a wealth of spectrophotometric information, making it readily applicable to a variety of astrophysical problems.}
{We used the \cmfgen\ code to calculate 606 NLTE, line-blanketed, expanding atmosphere models  using a comprehensive set of atomic data. An overall metallicity of 1/2 \zsun\ and 1/5 \zsun was adopted for the Large Magellanic Cloud (LMC) and Small Magellanic Cloud (SMC), respectively. We produced high-resolution spectra from 30~\AA\ to 3~$\mu$m for stars on the Main Sequence and slightly beyond.}
{We provide spectral energy distributions, normalized synthetic spectra, ionizing fluxes, and photometry in various bands: Johnson UBV, Cousins RI, Bessel JHK, selected wide \jwst\ filters, Gaia, and LSST ugrizy filters. For each of these filters, we compute bolometric corrections for all synthetic spectra and calibrations as a function of effective temperature.}
{All of our synthetic spectra are publicly available through the \pollux\  database, aiming to expedite multiwavelength analyses of massive stars in low metallicity environments.}

\keywords{Stars: massive -- Stars: atmospheres -- Stars: fundamental parameters -- Astronomical databases: miscellaneous}

\authorrunning{Marcolino et al.}
\titlerunning{Grids of CMFGEN spectra}

\maketitle

\section{Introduction}
\label{s_intro}

Our knowledge about massive stars has greatly improved since the first modern ultraviolet observations obtained detailed P-Cygni spectral profiles \citep{morton67,snow76,hutchings76,hutchings80}. It is now well established that the winds of O and B stars are mainly driven by momentum transfer of their intense radiation fields to metals present in the gas, i.e., they are line-driven winds \citep{lucysolomon70,cak,kud89,puls00}. Consequently, the strength of these winds, and thus the evolutionary paths of these stars from their birth to their final fate, hinges largely on the environment’s metallicity (Z). The quantitative aspects of this dependence have been the topic of numerous observational and theoretical studies and are still under thorough investigations \citep[e.g.,][]{leitherer92,bouret03,mokiem07,bjorklund21,marcolino22,hawcroft23, krticka24}.

Recently, the \ullyses\footnote{\url{https://ullyses.stsci.edu/}} and \xshootu\footnote{X-shooting ULLYSES - \url{https://massivestars.org/xshootu/}} observational programs provided an invaluable large ($\sim 240$ targets), homogeneous dataset of spectra of massive stars in the Large (LMC) and Small (SMC) Magellanic Clouds, alongside a few targets in the Local Group galaxies NGC 3109 and Sextans A \citep{vink23}. 
These galaxies all have sub-solar metallicity, ranging approximately from 1/2 to 1/10 Z$_{\odot}$. 
Within \ullyses, hundreds of Hubble Space Telescope (\hst) orbits were devoted to obtain high-resolution ultraviolet spectroscopy with the Cosmic Origins Spectrograph (COS) and Space Telescope Imaging Spectrograph (STIS). On the other hand, \xshootu\ focused on high-resolution optical-near IR observations of \ullyses\ targets with the \xshooter\ spectrograph on  the Very Large Telescope (\vlt). 
\textcolor{black}{Other initiatives, such as \bloem, which focuses on binaries with the \vlt\ \citep{shenar24}, and the 4MOST-\MONEMC\ survey \citep{cioni19}, will ultimately provide thousands of near-UV to near-IR spectra of massive stars in the Magellanic Clouds. Additionally, the James Webb Space Telescope (\jwst) has just opened a new window of possibilities for investigating the infrared emission characteristics of massive stars, an area that remains poorly explored \citep[see e.g.,][]{marcolino17}.}

Such multiwavelength observations of massive stars in various Local Group galaxies are crucial for understanding not only the quantitative effects of metallicity on stellar evolution, but also stellar populations in both nearby and distant galaxies. In this context, sophisticated, state-of-the-art atmosphere models are an indispensable component of this process.

Through the comparison of synthetic and high-resolution observed spectra, via diagnostic lines, essential physical parameters from both the photospheric and wind regions, as well as chemical abundances, can be obtained. These can then be directly compared with theoretical predictions, for example, from hydrodynamics or evolutionary calculations \citep{mokiem06,mokiem07,hunter08,ramirez13,markova14,martins15,holgado18,markova18,mahy20,bouret21,deburgos23,martins24}. 

Despite considerable progress in recent decades, the modeling of the expanding atmospheres of massive stars remains a big challenge. 
\textcolor{black}{Radiative transfer involving thousands of lines must be solved together with the statistical equilibrium equations (NLTE), as  
the LTE approximation is invalid.} A set of highly non-linear and coupled equations must be simultaneously solved. This task is complex and computationally expensive, but paramount for a proper interpretation of observed spectra \citep{hillier2012}. 
Consequently, the availability of pre-computed grids of synthetic spectra can greatly speed-up the analysis of large samples of stars. Such libraries of spectra can serve as input for techniques relying on machine learning methods that aim at providing fundamental parameters of OB stars. A large dataset of models can be also useful in other astrophysical contexts. Realistic ionizing sources 
are needed for photoionization codes (e.g., to investigate spectroscopy of H II regions), a library of synthetic spectra can aid stellar population synthesis analyses, and diverse model parameters can be used as well, as input in hydrodynamic calculations \citep[see e.g.,][]{vink01}.

\begin{figure*}[htbp]
  \centering   
  \includegraphics[width=\linewidth]{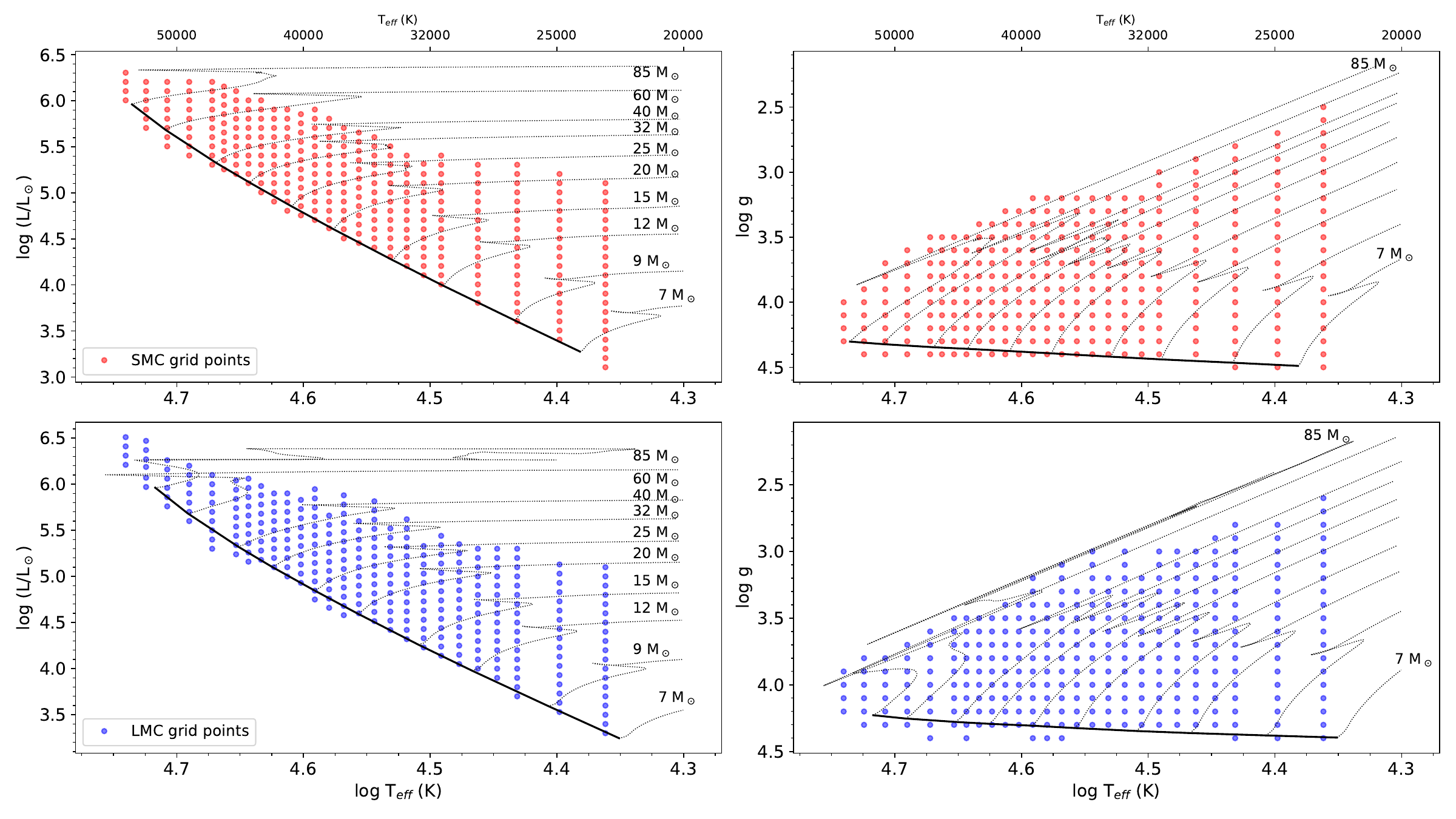} 
  \caption{SMC and LMC grid coverage in the Hertzsprung-Russell (left panels) and Kiel diagrams (right panels). Tracks from \citet{georgy13} and \citet{eggen21} are represented by dotted-lines. The thick solid line indicates the location of the ZAMS.}
  \label{fig:HRdiagram}
\end{figure*}

Grids of synthetic spectra for massive stars are publicly available. \citet{lh03} and \citet{lh07} present models for O and B stars respectively. The computations were performed with the code TLUSTY \citep{HL95_TLUSTY} and cover metallicities from twice solar to zero. The specificity of these grids is that they do not take spherical extension and stellar winds into account. 
The library of O stars synthetic spectra presented by \citet{bpass} overcome this limitation. The atmosphere code WM-BASIC \citep{pauldrach01} was used and the grid covers a wide range of metallicities \citep[see Sect. 2.3.2 and Table 2 of][]{bpass}. However WM-BASIC spectra do not include the physics required to produce detailed line profiles in the optical range.
More realistic synthetic spectra are presented in \citet{hainich19} and calculated with the \powr\ code \citep{graef02,sander15}. Solar, LMC and SMC metallicities have been chosen for these \powr\ grids that also extend to the range of parameters typical of Wolf-Rayet stars  \citep{todt15,sander12}.


The grids presented here are based on the \cmfgen\ code \citep[][]{hillier98} which is widely used for the spectroscopic analysis of massive stars, from the EUV to the infrared 
\citep[e.g.,][]{hillier03,martins05,marcolino09,martins15,marcolino17,almeida19,bouret21}.
Libraries of synthetic spectra at solar and sub-solar metallicity are already available through the \pollux\ database\footnote{\url{https://pollux.oreme.org/}}. \citet{zsargo20} also provide large model grids with special focus on wind properties. However these libraries do not follow the classical parameter coverage of grids, i.e. do not have a regular sampling in effective temperature and surface gravity. Rather, they correspond to spectra computed along evolutionary tracks \citep{mp17,mp21,mp22,zsargo20}. Here we complement these data by proper grids of models at sub-solar metallicity. We generated a total of 606 models based on the \cmfgen\ code, from which we produced high-resolution spectra from the EUV to mid-IR. In addition to synthetic spectra, we provide synthetic photometry, bolometric corrections, and ionizing fluxes. We make all spectra publicly available at the \pollux\ database.


The paper is organized as follows. In Sect. \ref{s_grid} we describe in detail how we constructed our grids. Sect. \ref{s_data} presents the spectrophotometric data we generated. We discuss the main characteristics, magnitudes and bolometric corrections in different filters, as well as the ionizing radiation in this section. Finally, we summarize the main conclusions of our paper in Sect. \ref{s_conclusions}.

\section{Building the grids}
\label{s_grid}

\subsection{CMFGEN: Main ingredients}
\label{ss_cmf}

The grids were constructed using the \cmfgen\ code by \cite{hillier98}. 
Initially, \cmfgen\ was designed to compute spherically symmetric 
expanding model atmospheres in NLTE to simulate massive, hot stars with powerful, optically thick winds. It has been since used in a great variety of cases, from OB stars with weak winds, in the Galaxy or in the Magellanic Clouds (MCs), to all types of supernovae (core-collapse and thermonuclear), Wolf-Rayet stars, Luminous Blue Variables, and central stars of planetary nebulae
\citep{aadland22,blondin15,dessart05,dessart15,dessart23,groh09,groh11,hillier87,hillier99,hillier20,marcolino07,marcolino07b,martins04,morris17,neugent17,teodoro20}.

\cmfgen\ solves the radiative transfer and statistical equilibrium equations simultaneously within the co-moving frame (CMF) framework, including a treatment of line-blanketing using a super-level approach \cite[][]{hillier98, hillier99}. The super-levels formalism allows for the inclusion of multiple energy levels from various ion species in the model atmosphere calculations. More details about this are given in Sect. \ref{ss_atom} (see also Table \ref{tab_lev}). 

The wind structure is treated in the following way: the radiative acceleration is computed from the solution of the level populations, and is used to determine in an iterative way the density structure of the inner atmosphere in a quasi-hydrostatic region. The resulting velocity law is smoothly connected to a $\beta$ velocity law in the wind \cite[see e.g.,][]{bouret12}. The wind mass-loss rate, density, and velocity are related through the continuity equation. 

To address the unstable nature of radiatively driven winds, \cmfgen\ can incorporate the effects of X-rays and wind clumping. The optically-thin clumping approximation is used \cite[see e.g.,][]{bouret05}, but the code was recently modified to include shells, handling clumps with arbitrary optical depths \citep{flores21,flores23}. However, to facilitate the grid construction, we decided to not include clumping nor X-rays in the models. Both ingredients can be easily switched on when starting the analysis of an observed spectrum. 

Once the atmosphere model is converged, a formal solution of the radiative transfer equation is computed in the observer's frame, providing a synthetic spectrum for comparison with observations. In this step, \cmfgen\ takes into account a radially increasing microturbulent velocity, starting at about 10 \kms\ at the photosphere \citep[see e.g.,][for the adopted formula]{hillier03} and extending up to $\sim$10\% of the terminal velocity in the outer atmosphere. Detailed line profiles and broadening mechanisms are taken into account. 

\subsection{Grid setup}
\label{ss_gridcoverage}

The parameter space encompassed by our grids is anticipated to accurately depict the characteristics of actual OB stars in the Magellanic Clouds, aligning with recent high-resolution ultraviolet and optical observations conducted under the \ullyses\ and \xshootu\ programs. 

We built two grids of models, both assuming 
a solar helium abundance, He/H = 0.085 (by number). All other chemical abundances are scaled from the solar values \citep{asplund09} assuming a metallicity of 1/2 and 1/5 \zsun\ for the LMC and SMC, respectively. 

\textcolor{black}{The effective temperature (\teff) range from 23 kK to 55 kK.
Most of the grid was computed with a 1 kK step, but nearing both \teff\ ends, a coarser step of 2kK was adopted\footnote{\textcolor{black}{Less \ullyses\ targets are expected towards these effective  temperatures \citep[see e.g., Fig. 2 of][]{vink23}.}}. In total, we have 25 temperatures for the SMC and 26 temperatures for the LMC}. For each \teff, we calculate models for a series of surface gravities (\logg's), adopting a 0.1 dex step. The first value of \logg\ and \logL\ of the grid are the rounded value of \logg\ on the zero-age main-sequence (ZAMS) (1st point) for Geneva models (from \citealt{georgy13} and \citealt{eggen21}). \textcolor{black}{Then, as we decrease \logg\ by 0.1 dex, we increase \logL\ by 0.1 dex. For example, for a first model with \logL\ = 5.0 for \logg\ = 4.4, the second model in the series will have  \logL\ = 5.1 for \logg\ = 4.3, and so on. This approach is equivalent to leaving the mass unchanged for a sequence of models with fixed \teff}. The aim of this empirical sampling in \logg\ (or equivalently \logL) is to sample the main-sequence (MS) and a little beyond for a given \teff\ (i.e. to make a ''vertical line'' covering enough of the MS in the HRD).

\begin{table}[ht]
\begin{center}
\caption{Ions, number of super-levels and number of levels included in the model calculations. Ions with an asterisk symbol are only used in low or high \teff\ models.} \label{tab_lev}
\begin{tabular}{lcc}
\hline
Ion  & \# super-levels & \# levels\\    
\hline
\ion{H}{i}      &  30  & 30   \\     
\ion{He}{i}     &  69  & 69   \\    
\ion{He}{ii}    &  30  & 30   \\
\ion{C}{ii*}  	&  92  & 322  \\     
\ion{C}{iii}    &  99  & 243  \\    
\ion{C}{iv}     &  64  & 64   \\
\ion{N}{ii*}  	&  59  & 105  \\
\ion{N}{iii}    &  57  & 287  \\   
\ion{N}{iv}     &  44  & 70   \\    
\ion{N}{v}      &  41  & 49   \\
\ion{O}{ii*}  	&  155 & 274  \\
\ion{O}{iii}    &  36  & 104  \\    
\ion{O}{iv}     &  30  & 64   \\    
\ion{O}{v}      &  32  & 56   \\
\ion{O}{vi*}	&  9   & 15   \\
\ion{Ne}{ii}    &  14  & 48   \\    
\ion{Ne}{iii}   &  23  & 71   \\    
\ion{Ne}{iv}    &  17  & 52   \\    
\ion{Ne}{v}     &  37  & 166  \\    
\ion{Mg}{ii}    &  36  & 44   \\
\ion{Si}{ii*} 	&  27  & 53   \\
\ion{Si}{iii}   &  50  & 50   \\    
\ion{Si}{iv}    &  66  & 66   \\    
\ion{S}{iii}    &  39  & 78   \\    
\ion{S}{iv}     &  40  & 108  \\    
\ion{S}{v}      &  37  & 144  \\
\ion{S}{vi*}    &  28  & 58   \\
\ion{Ar}{iii}   &  24  & 138  \\   
\ion{Ar}{iv}    &  30  & 102  \\   
\ion{Ar}{v}     &  14  & 29   \\
\ion{Ar}{vi*}   &  21  & 81   \\
\ion{Ca}{iii}   &  29  & 88   \\   
\ion{Ca}{iv}    &  19  & 72   \\
\ion{Ca}{v*} 	&  10  & 613  \\
\ion{Fe}{ii*} 	&  24  & 295  \\
\ion{Fe}{iii}   &  65  & 607  \\   
\ion{Fe}{iv}    &  100 & 1000 \\ 
\ion{Fe}{v}     &  139 & 1000 \\ 
\ion{Fe}{vi}    &  59  & 1000 \\
\ion{Fe}{vii*}  &  41  & 252  \\
\ion{Ni}{ii}*   &  27  & 158  \\
\ion{Ni}{iii}   &  24  & 150  \\   
\ion{Ni}{iv}    &  36  & 200  \\   
\ion{Ni}{v}     &  46  & 183  \\   
\ion{Ni}{vi}    &  40  & 182  \\   
\ion{Ni}{vii*}  &  37  & 308  \\
\hline                     
\end{tabular}
\end{center}
\end{table}

We assumed a fixed micro turbulence velocity of 10 \kms. The wind acceleration parameter is set to $\beta$ = 1.0 and the mass-loss rates are computed using the \citet{vink01} formula for the corresponding metallicity, assuming that the terminal velocity is proportional to the escape velocity as 
$v_{\infty}/v_{esc} = 3.0$, to be consistent with \citet{mp21} and references therein. \textcolor{black}{Other mass-loss recipes exist. However, there is no consensus on which one is preferable. For example, the predictions for the mass-loss rate dependence with luminosity (and metallicity) scatter \citep[see e.g., fig. 5 of][]{bjorklund21}. Vink's recipe remains the most widely used in evolutionary calculations. As previously mentioned, our initial model grid points were anchored to ZAMS track points that utilize Vink.}

We present the LMC and SMC grid coverage in the Hertzsprung-Russell and Kiel diagrams in Fig. \ref{fig:HRdiagram}. Our points, each corresponding to an atmosphere model, are shown along with theoretical tracks from \citet{georgy13} and \citet{eggen21}, that include rotation. We stress that our models are not tailored to follow the tracks and the latter are shown for illustration purpose. A large interval of initial mass is covered for each metallicity, representing both O and B stars. For several masses, the points sample well the whole Main Sequence. 

\textcolor{black}{We note that both grids may be updated in the future, for instance, by filling gaps in temperature, adding more ions, and extending the mass range. For instance, it is certainly worth exploring models above 100 M$_\odot$. Despite being extremely rare, very massive stars (VMS) have been observed, modeled, and have been shown to be crucial for population synthesis studies in nearby galaxies \citep[see e.g.,][and references therein]{martins23}. Grid points for VMS are beyond the scope of the present paper, but models and spectra along evolutionary tracks have been computed in the literature \citep[see e.g.,][]{mp22}.}

\subsection{Atomic data}
\label{ss_atom}

OB stars exhibit a rich ultraviolet spectrum characterized by 
numerous metallic absorption lines \citep[e.g.,][]{pauldrach1993}. 
Primarily originating from the photosphere (e.g., iron forest), these 
lines are often accompanied by prominent wind P-Cygni profiles (e.g., \ion{C}{iv} $\lambda$ 1549, \ion{N}{v} $\lambda$ 1240, \ion{O}{v} $\lambda$ 1371) that are important diagnostics of the mass-loss rate and the outflow velocity structure. Conversely, their optical spectra are dominated mainly by strong hydrogen and helium lines. These are in general observed in absorption, but some may be in emission depending on the wind strength (e.g, H$\alpha$ in supergiants). 

Spectral analyses that neglect elements other than H, He, and CNO, inaccurately infer important physical parameters. The presence of numerous metal lines blocks and scatter back radiation to inner parts of the atmosphere, resulting in a back-warming, implying a change in the ionization structure of a model in comparison with, for example, a metal free one \citep[see e.g.,][]{martins02,lh03,lh07}. In fact, models with the same effective temperature but different metallicities can impact \ion{He}{i-ii} line strengths \citep{martins02}.

In our grids, we choose to incorporate as many elements and ions as possible, having 
a compromise with an appropriate amount of blanketing (e.g., CNO+iron elements ions) and a manageable model, i.e., not too large and very computationally expensive. We present the dataset used in Table \ref{tab_lev}. 
A total of 2046 super-levels were formed from a total of 9178 levels \cite[for more details, see][]{hillier98}. In practice, they correspond up to hundreds of thousands transitions to be taken care of in the radiative transfer calculation.

\subsection{Availability and output products}
\label{ss_pollux}

The 606 spectra of our grids are made publicly available at the POLLUX database\footnote{\url{https://pollux.oreme.org}}  \citep[][]{pollux}). The entire SED can be downloaded, as well as cuts of the UV, visible and infrared spectrum. A quick look tool with zoom facilities allows inspections of specific features prior to download. For each file a header describes the atmosphere model parameters and the parameters of the formal solution of the radiative transfer that lead to the synthetic spectra. The non-normalized and normalized spectra can be directly compared to observations, after including the necessary natural and instrumental broadening effects. We recall that \cmfgen\ calculates a spectrum in the reference frame of an observer at 1 kpc, by default. Thus, a proper distance scaling and addition of ISM extinction effects must be taken into account when handling the non-normalized data. The atmosphere models are available upon request. They can be used to construct new ones or to obtain several important physical properties of the atmosphere -- for instance, ionization structures, level populations -- which are relevant for various purposes.

\section{Spectrophotometric data}
\label{s_data}

\subsection{Synthetic high-resolution spectra}

\begin{figure*}
  \centering
  \includegraphics[width=\textwidth]{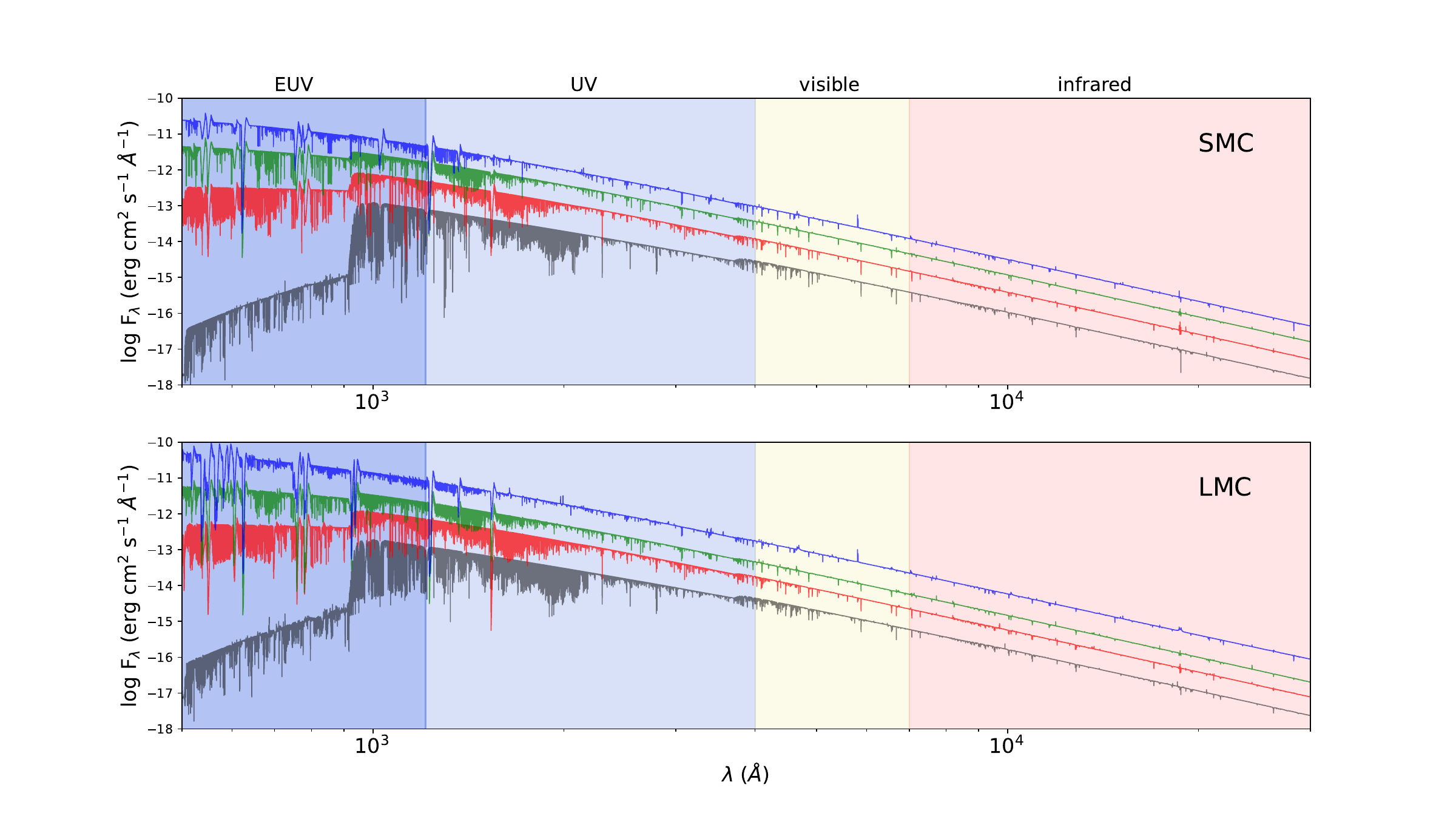} 
  \caption{Illustrative spectra from our grid of models representing LMC and SMC OB stars. The displayed wavelength range spans from 500\AA\ to 30000\AA. All models have a fixed log $g = 4.0$ and \teff\ = 25, 35, 45 and 55kK (bottom to top). Roughly, they correspond to an early B star, late, mid and early O star, respectively. Background colors correspond to different wavelength ranges as indicated by the top labels.}
  \label{fig:panview}
\end{figure*}

\begin{figure}[!h]
  \centering
  \includegraphics[width=\linewidth]{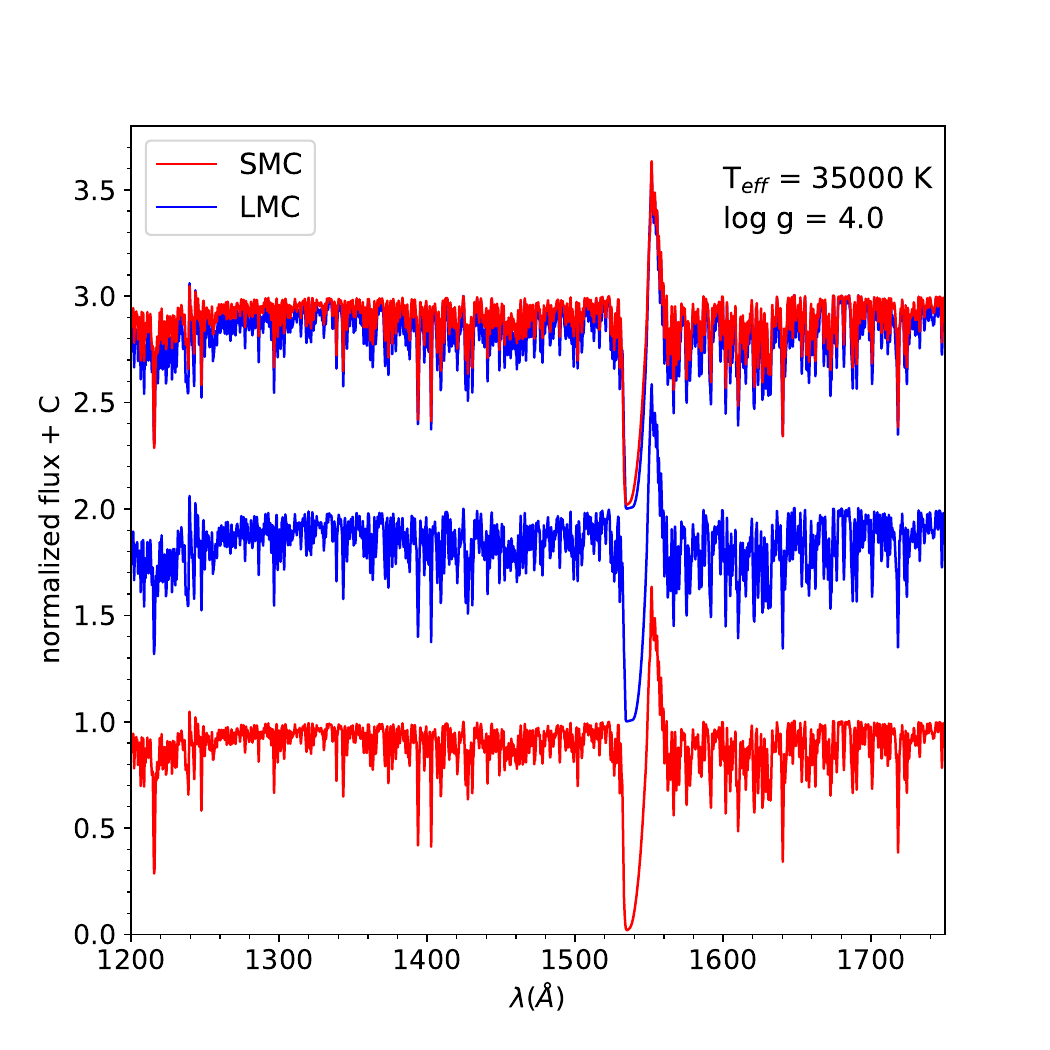} 
  \caption{Zoom on the far-UV spectrum of two models with similar \teff\ and \logg\ but different metallicities (hence wind parameters). The blue (red) spectrum is the LMC (SMC) model. The spectra were convolved with a rotational profile of 100 \kms\ and a resolving power R = 50000.}
  \label{fig:UVZ}
\end{figure}

\begin{figure}[!h]
  \centering
  \includegraphics[width=\linewidth]{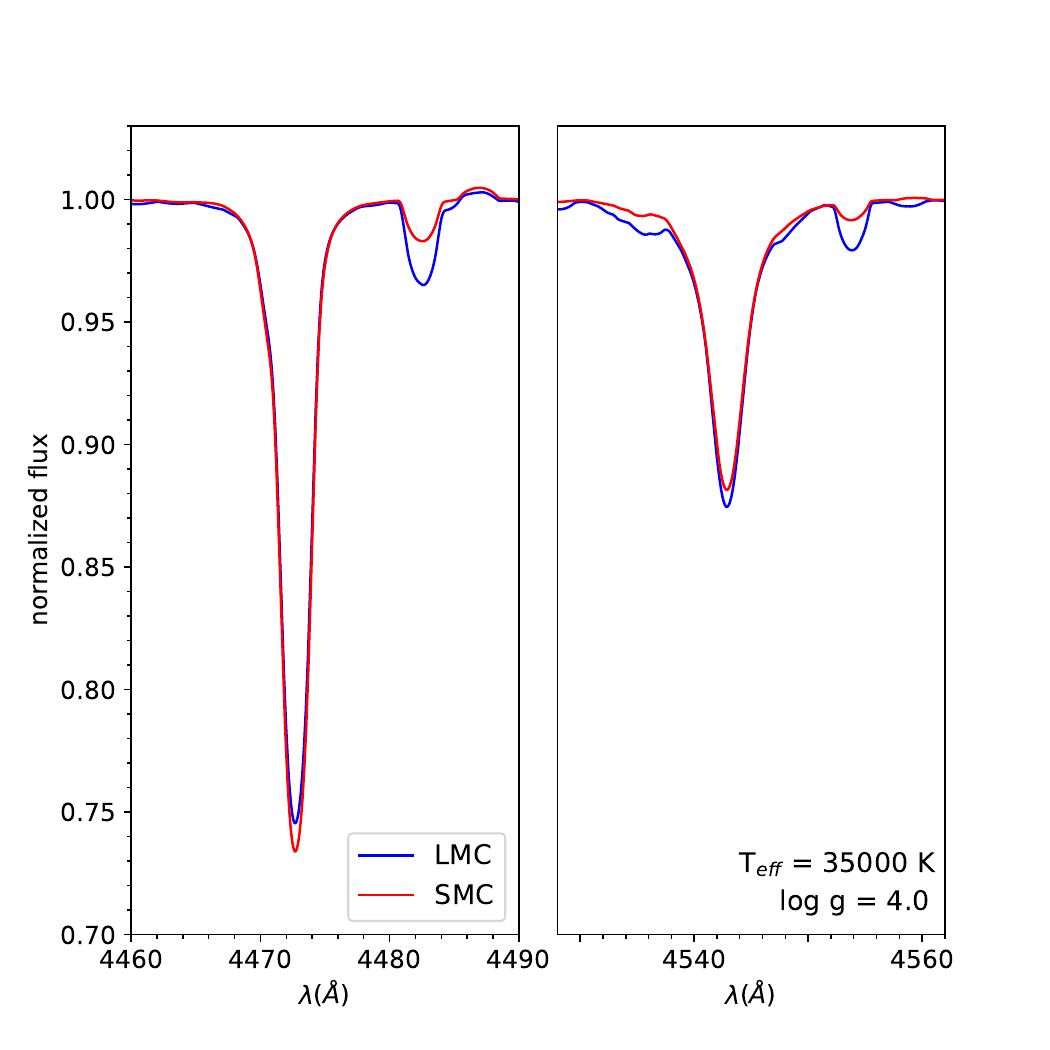} 
  \caption{Same as for Fig.~\ref{fig:UVZ}, but for the optical lines \ion{He}{i}~4471 (left panel) and \ion{He}{ii}~4542 (right panel).}
  \label{fig:optZ}
\end{figure}

We generated high-resolution spectra spanning from extreme ultraviolet (EUV) to mid-infrared (mid-IR) for all models within the LMC and SMC grids. These spectra can yield several relevant physical parameters, some of which will be discussed in the next sections.

In Fig. \ref{fig:panview} we present a small set of spectra to illustrate the quality of the models, over a large wavelength interval. 
They have log $g = 4.0$ and the effective temperature \teff\ is 25, 35, 45 and 55$k$K, from bottom to top. \textcolor{black}{The fluxes are absolute and were scaled taking into account the LMC and SMC distances \citep[][]{pietrzynski2019,graczyk2020,vink23}. No reddening or extinction is simulated.}

The highest flux and density of wind and photospheric lines in the EUV/UV is apparent, a common characteristic of OB stars. Overall, the LMC models present more absorption features than the SMC ones, reflecting the higher metallicity. Hundreds of lines from the EUV to the infrared can be seen and readily compared with real observed spectra, after the appropriate rotational, macroturbulence, and/or instrumental convolution. 

Figure \ref{fig:UVZ} shows how metallicity directly impacts the far-UV spectra for models with similar fundamental parameters. The iron line forests that make most of the absorption features in that figure are weaker at lower Z. Helium lines are also impacted by the indirect metallicity effects of line-blocking and back-warming. We see in Fig. \ref{fig:optZ} that even if the He/H ratio is unchanged between both grids, the \ion{He}{i}~4471 line is weaker at higher Z, while the \ion{He}{ii}~4542 line is stronger. This is due to the different temperature structure at different metallicity (see \citealt{martins02,repolust04} for details). These lines are key features to constrain \teff\ in spectroscopic analysis.

In the Appendix, we show additional spectra from our grid models across selected wavelength intervals, highlighting their practical significance. Figure \ref{fig:uvspectra} focus on the UV/EUV range, its richness and changes due to the metallicity. Figure \ref{fig:logteff-lines} illustrates the effects of 
\teff\, and \logg\, variations on selected hydrogen and helium lines. In Fig. 
\ref{fig:ir-lines} we provide examples of infrared spectra.

\subsection{Photometry}
\label{ss_photo}

The spectra of our model grids were computed over a large wavelength interval, thus allowing the calculation of magnitudes in various bands of interest. Synthetic photometry is important for several reasons. Colors can be computed, as well as absolute magnitudes, which can thus be used to compute bolometric corrections (BC). The variation of the BC with the effective temperature can then be explored. Bolometric corrections are useful to estimate stellar luminosities from photometric surveys. 

To compute the absolute magnitudes in each band, we proceeded as follows: each model spectrum of our LMC and SMC grids was first re-scaled from 1 kpc -- the \cmfgen\ output default -- to 10 pc. Then, we used each SED as input in the Python package PYPHOT, created by Morgan Fouesneau\footnote{\url{https://mfouesneau.github.io/pyphot/}}. The flux points and then magnitudes are computed according to the following formula:

\[ M_\lambda = -2.5 \times log \int F_\lambda B_\lambda d\lambda + const.        \]

\noindent where $B_\lambda$ is the filter passband. With the absolute magnitudes, the bolometric corrections follow straightforwardly from:

\[  BC_{\lambda} = M_{\odot}^{bol} - M_{\lambda} - 2.5 \times log \frac{L}{L_\odot}   \]

\noindent where $M_{\odot}^{bol}$ is the solar bolometric magnitude, equals to 4.75 (IAU 1999), and $L/L_\odot$ is the model luminosity in solar units. 

\begin{figure}[htbp]
  \centering
  \includegraphics[width=\linewidth]{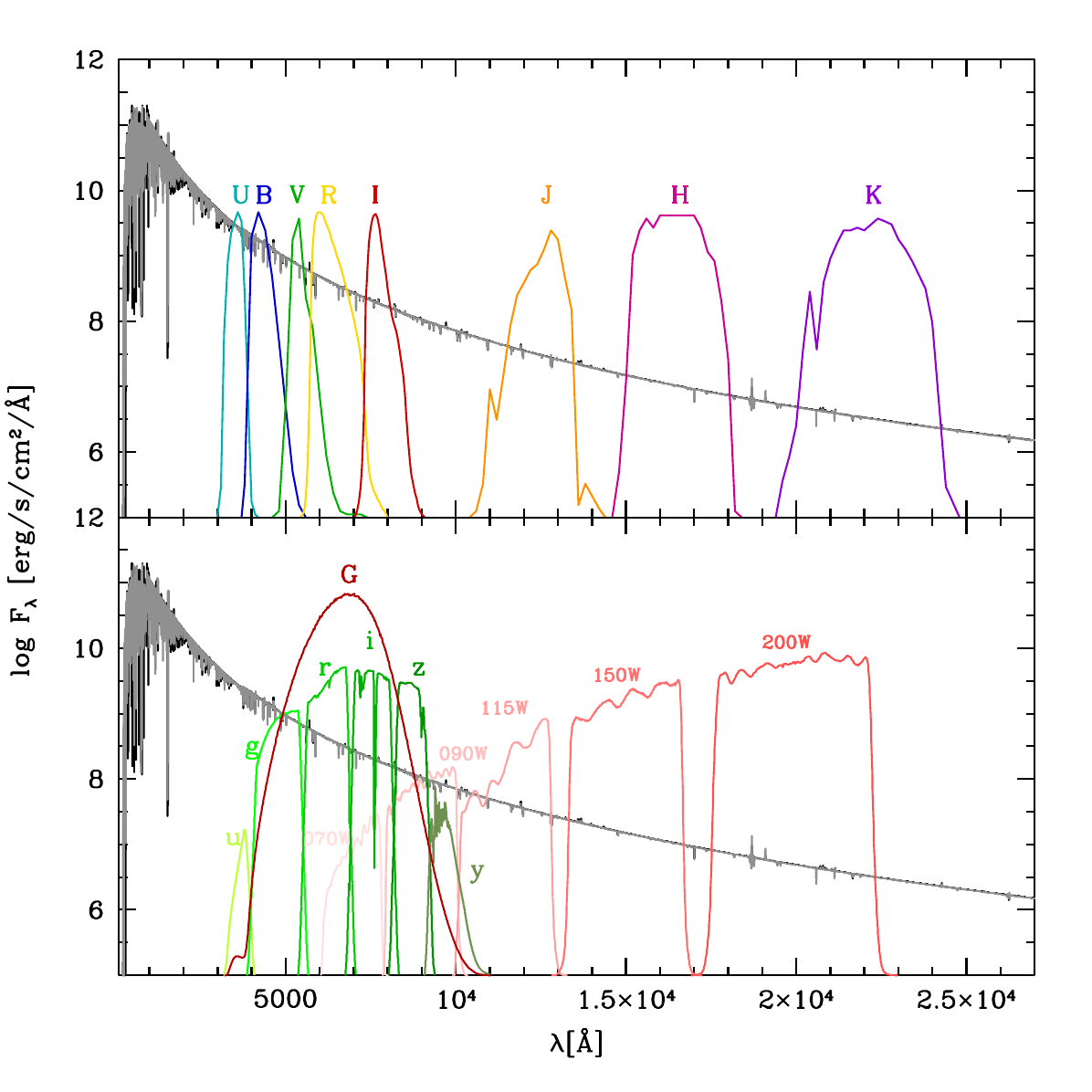} 
  \caption{Illustration of the position of the selected filters for computation of photometry. The top panel shows the Johnson UBV, Cousins RI, and Bessel JHK filters. The bottom panel shows the Gaia G band (dark red), LSST (levels of green) and JWST (levels of pink) filters. GAIA Gbp and Grp filters were omitted, for clarity. Two models at LMC (black) and SMC (grey) metallicity are shown. They both have \teff\ =~37000~K and \logg\ =~3.6.}
  \label{fig:filters}
\end{figure}

We used the Johnson UBV, Cousins RI, and Bessel JHK photometric bands. We also computed magnitudes in some filters from  \jwst, namely, F070W, F090W, F115W, F150W, and F200W\footnote{The number is the central wavelength of the filter in 10$^{-2} \mu$m units and W stands for wide. For more details, see the \href{https://jwst-docs.stsci.edu/jwst-near-infrared-camera/nircam-instrumentation/nircam-filters}{JWST User Documentation}.}. This is the first time that \jwst\ absolute magnitudes of models representing OB stars are presented. \textcolor{black}{For the benefit of the Gaia mission we provide G, Gbp, and Grp magnitudes (DR3)}. Finally, in preparation of the Vera Rubin survey, we calculated magnitudes in the six Legacy Survey of Space and Time (LSST) filters (u, g, r, i, z, and y).
Details about all filters used – for example, centers, widths, transmission curves and zero points – can be found at the Spanish Virtual Observatory website (Rodrigo, Solano \& Bayo 2012; Rodrigo \& Solano 2020)\footnote{\url{http://svo2.cab.inta-csic.es/svo/theory/fps3/}}. Figure~\ref{fig:filters} shows the transmission curves of all filters considered and their relative wavelength position compared to typical SEDs of our grid models. The complete dataset, with all the absolute magnitudes and bolometric corrections in each photometric band, will be available in the VizieR catalog (CDS, Strasbourg).

\begin{figure*}[htbp]
  \centering
  \includegraphics[width=\linewidth]{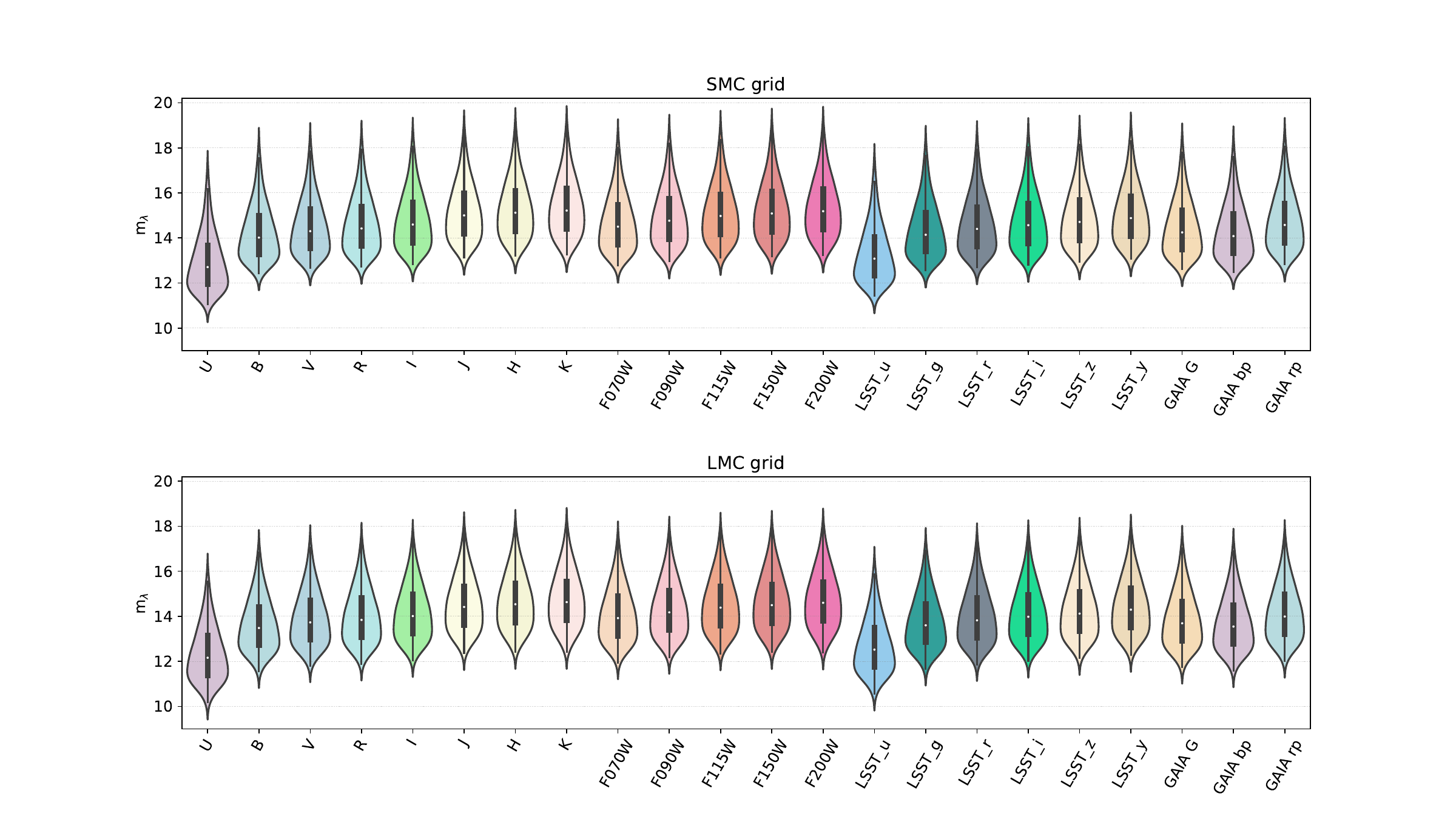} 
  \caption{Synthetic photometry of OB stars in the LMC and SMC, as calculated from our models. We show the distribution of magnitudes in several bands through a violin statistical plot. The adopted distances for the LMC and SMC follow the distance moduli of the LMC and SMC \citep[][]{pietrzynski2019,graczyk2020,vink23}. ISM effects are not taken into account (see text for more details). }
  \label{fig:violin}
\end{figure*}

\begin{table}[h!]
\centering
\caption{Fit coefficients for the BC$_\lambda$--\teff\ relation at SMC metallicity.}
\label{tab:coeffsSMC}
\begin{tabular}{l|ccc}
\hline
\hline
\textbf{Band} & \textbf{intercept} & \textbf{slope} & \textbf{rms} \\
\hline
U & $23.32 \pm 0.36$ & $-5.53 \pm 0.08$ & 0.14 \\
B & $25.28 \pm 0.28$ & $-6.25 \pm 0.06$ & 0.11 \\
V & $25.75 \pm 0.22$ & $-6.41 \pm 0.05$ & 0.08 \\
R & $25.87 \pm 0.19$ & $-6.46 \pm 0.04$ & 0.07 \\
I & $26.03 \pm 0.17$ & $-6.53 \pm 0.04$ & 0.06 \\
J & $26.55 \pm 0.14$ & $-6.73 \pm 0.03$ & 0.05 \\
H & $26.62 \pm 0.16$ & $-6.77 \pm 0.03$ & 0.06 \\
K & $26.67 \pm 0.18$ & $-6.80 \pm 0.04$ & 0.07 \\
F070W & $25.92 \pm 0.18$ & $-6.49 \pm 0.04$ & 0.07 \\
F090W & $26.28 \pm 0.15$ & $-6.62 \pm 0.03$ & 0.06 \\
F115W & $26.57 \pm 0.14$ & $-6.73 \pm 0.03$ & 0.05 \\
F150W & $26.58 \pm 0.15$ & $-6.73 \pm 0.03$ & 0.06 \\
F200W & $26.65 \pm 0.17$ & $-6.79 \pm 0.04$ & 0.07 \\
LSST u & $23.79 \pm 0.34$ & $-5.72 \pm 0.08$ & 0.13 \\
LSST g & $25.47 \pm 0.26$ & $-6.31 \pm 0.06$ & 0.10 \\
LSST r & $25.85 \pm 0.20$ & $-6.45 \pm 0.04$ & 0.07 \\
LSST i & $25.98 \pm 0.17$ & $-6.51 \pm 0.04$ & 0.06 \\
LSST z & $26.20 \pm 0.15$ & $-6.59 \pm 0.03$ & 0.06 \\
LSST y & $26.50 \pm 0.14$ & $-6.69 \pm 0.03$ & 0.05 \\
Gaia G & $25.50 \pm 0.23$ & $-6.35 \pm 0.05$ & 0.09 \\
Gaia Gbp & $25.23 \pm 0.26$ & $-6.25 \pm 0.06$ & 0.10 \\ 
Gaia Grp & $26.01 \pm 0.17$ & $-6.52 \pm 0.04$ & 0.06 \\ 
\hline
\end{tabular}
\end{table}

\begin{figure*}[htbp]
  \centering
  \includegraphics[width=\linewidth]{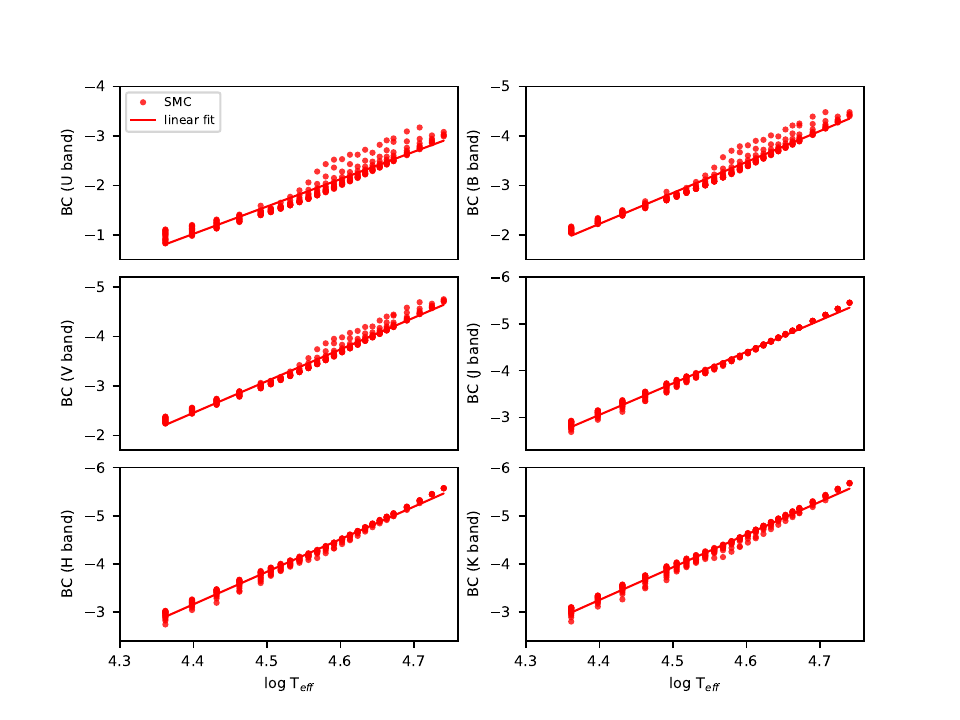} 
  \caption{Bolometric corrections versus the effective temperature in selected  photometric bands (see text for details). Our SMC data is indicated by solid  circles and the corresponding linear fit by a solid line.}
  \label{fig:BCteff}
\end{figure*}

\begin{table}[h!]
\centering
\caption{Fit coefficients for the BC$_\lambda$--\teff\ relation at LMC metallicity.}
\label{tab:coeffsLMC}
\begin{tabular}{l|ccc}
\hline
\hline
\textbf{Band} & \textbf{intercept} & \textbf{slope} & \textbf{rms} \\
\hline
U & $23.38 \pm 0.34$ & $-5.54 \pm 0.07$ & 0.12 \\
B & $25.22 \pm 0.26$ & $-6.23 \pm 0.06$ & 0.10 \\
V & $25.76 \pm 0.21$ & $-6.41 \pm 0.05$ & 0.08 \\
R & $25.81 \pm 0.18$ & $-6.44 \pm 0.04$ & 0.07 \\
I & $26.01 \pm 0.17$ & $-6.52 \pm 0.04$ & 0.06 \\
J & $26.50 \pm 0.15$ & $-6.71 \pm 0.03$ & 0.05 \\
H & $26.50 \pm 0.17$ & $-6.74 \pm 0.04$ & 0.06 \\
K & $26.41 \pm 0.20$ & $-6.74 \pm 0.04$ & 0.07 \\
F070W & $25.92 \pm 0.17$ & $-6.48 \pm 0.04$ & 0.06 \\
F090W & $26.27 \pm 0.15$ & $-6.61 \pm 0.03$ & 0.06 \\
F115W & $26.49 \pm 0.14$ & $-6.71 \pm 0.03$ & 0.05 \\
F150W & $26.50 \pm 0.16$ & $-6.73 \pm 0.03$ & 0.06 \\
F200W & $26.41 \pm 0.20$ & $-6.73 \pm 0.04$ & 0.07 \\
LSST u & $23.83 \pm 0.32$ & $-5.72 \pm 0.07$ & 0.12 \\
LSST g & $25.43 \pm 0.24$ & $-6.30 \pm 0.05$ & 0.09 \\
LSST r & $25.81 \pm 0.19$ & $-6.44 \pm 0.04$ & 0.07  \\
LSST i & $25.96 \pm 0.17$ & $-6.50 \pm 0.04$ & 0.06 \\
LSST z & $26.18 \pm 0.16$ & $-6.58 \pm 0.03$ & 0.06 \\
LSST y & $26.45 \pm 0.14$ & $-6.68 \pm 0.03$ & 0.05 \\
Gaia G & $25.48 \pm 0.22$ & $-6.33 \pm 0.05$ & 0.08 \\
Gaia Gbp & $25.18 \pm 0.24$ & $-6.24 \pm 0.05$ & 0.09 \\
Gaia Grp & $26.02 \pm 0.16$ & $-6.52 \pm 0.04$ & 0.06 \\
\hline
\end{tabular}
\end{table}

Figure \ref{fig:violin} shows the distribution of apparent magnitude values (m$_\lambda$’s) inferred from all synthetic spectra, from both the LMC and SMC grids\footnote{The corresponding plot with absolute magnitudes is obviously similar, the values are merely shifted in the vertical axis by the respectives distance moduli.}. A total of 13332 magnitudes are displayed using violin plots. Their shapes illustrate the data density in each band. Medians are indicated by white dots. Thick black bars represent the interquartile ranges, giving an indication of the spread of 50\% of the data. Thin lines within each violin represent the range of values.

The classical UBVRIJHK magnitudes are followed by JWST, LSST, and GAIA magnitudes. The apparent magnitudes were computed from the equation $m_\lambda = M_\lambda - 5 + 5 \times \log d$, using the distances of the LMC and SMC. ISM effects are not taken into account. 

Overall, the distributions increase in magnitude from blue to redder filters, as expected. Moreover, they possess similar shapes, without anomalies (e.g., skewness, multiple peaks), indicating that the models and flux calculations were carried out properly. Fig. \ref{fig:violin} also provides a graphical visualization of apparent magnitudes that massive OB stars in the Magellanic Clouds may have in each band, in case of minimal or no extinction line of sights. However, the distributions shown serve as indicators rather than definitive representations.
For real objects, the distributions are expected to be shifted differentially and upwards (by m$_\lambda + A_\lambda$). The position of the medians could also be affected, since some spectral types are more abundant than 
others (e.g., through the IMF and observational biases).


\begin{figure*}[htbp]
  \centering
  \includegraphics[width=\linewidth]{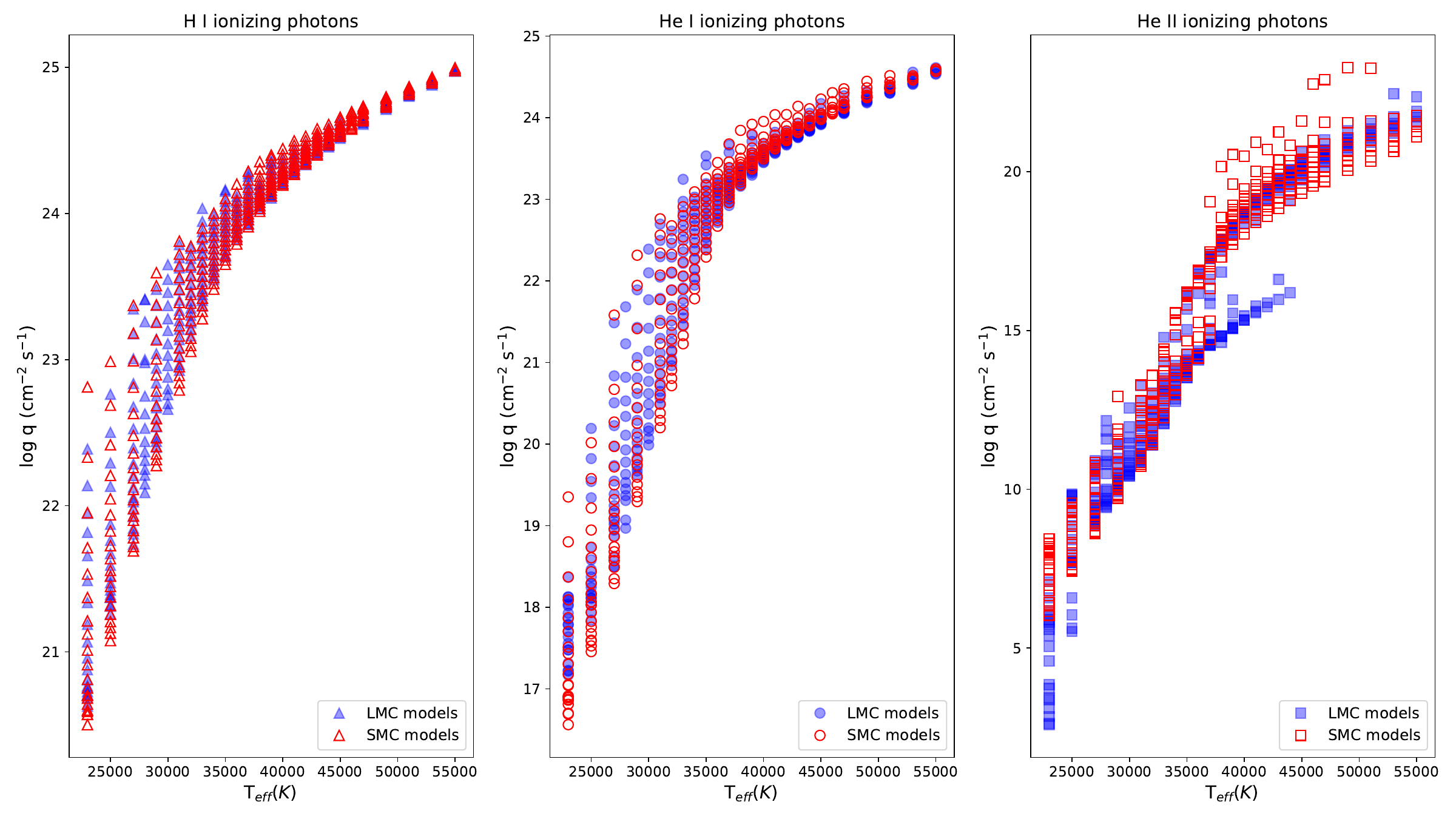} 
  \caption{Logarithm of the number of ionizing photons per unit surface area (log q) as a function of the effective temperature from our grid models. From left to right we show log q above the ionization threshold for \ion{H}{i}, \ion{He}{I}, and \ion{He}{II}. Red (blue) points represent the SMC (LMC) grid.}
  \label{fig:logq}
\end{figure*}


Figure \ref{fig:BCteff} shows the bolometric corrections in the most used, classical UBVJHK bands, as a function of the effective temperature. Only the SMC points are displayed. Nevertheless, we provide in Tables \ref{tab:coeffsSMC} and \ref{tab:coeffsLMC} the coefficients of the linear relations for all bands, for the SMC and LMC, respectively. 

The absolute values of the bolometric corrections increase with temperature, as expected. Indeed, using the Wien maximum as a first approximation, the peak of the spectral energy distribution occurs at shorter wavelengths for higher effective temperatures. For instance, at approximately 1159\AA\ for log T$_{\rm eff}$ $\sim$ 4.4 ($\sim$ 25000K) and at $\sim$ 580\AA, for log T$_{\rm eff}$ $\sim$ 4.7 ($\sim$ 50000K). Therefore, this peak shifts away from the wavelength centers of the different filters we consider as we go towards high effective temperatures, meaning higher bolometric corrections.

The effective temperature is not the only parameter that influences the bolometric corrections. For the U, B and V bands at log T$_{\rm eff}$ $\sim$ 4.6, for example, there is a spread of BC values. This is a natural consequence of the way our grids were constructed. For a specific temperature, we explored different values of surface gravities and luminosities and thus mass-loss rates, affecting the photometry (see Sect. \ref{s_grid}). 

\citet{martins06} provide a calibration of bolometric correction versus \teff\ for Galactic stars. It is tempting to compare directly the present results to this calibration to investigate metallicity effects. However this task is rather complex for the following reasons. First, \citet{martins06} used generic Johnson filters as we do in this work for UBV bands, but the exact shape of the transmission curve and the associated zero points are not strictly the same. This leads to systematic variations in photometry. Second, the present models are more complete than those of \citet{martins06}. In particular, we include more elements, levels, and lines. The resulting line-blanketing effects that affect the shape of the SED are thus not of the same order in the two studies. This introduces another source of potential difference in synthetic photometry. Third, we assume that the relations between bolometric corrections and $\log$(\teff) are linear. Inspection of Figs.~\ref{fig:BCteff} reveals that this is a very good assumption for some temperature ranges, but only approximate for the full interval. Depending on the range of \teff\ considered, the coefficients of the linear regressions vary. As an example, restricting the \teff\ range to 30000-45000~K leads to $BC(V)=28.86-7.07$ $\times$ $\log$ (\teff). This is significantly steeper than the relation reported in Table~\ref{tab:coeffsLMC}. Since the relations of \citet{martins06} are based on models that cover a narrower range of \teff\ than the present study, a direct comparison between their and our relations cannot provide any insight into the metallicity dependence of bolometric corrections.

However, what is feasible is a comparison between the SMC and LMC relations of the present study, since they rely on the same photometric set-up and on grids that cover the same parameter space. Tables~\ref{tab:coeffsSMC} and \ref{tab:coeffsLMC} show that for a given \teff\ bolometric corrections are very similar, hinting at the absence of metallicity effect between the SMC and the LMC. \textcolor{black}{In fact, the slopes and intercepts in each band are compatible between the LMC and SMC cases within the uncertainties.} 

\textcolor{black}{We would like to remind readers that line-blanketing and thus back-warming, as well as flux redistribution, change the SED of models with different metallicities. However, our results indicate that to possibly observe significant differences in the BCs and thus in the calibrations, one has to consider a larger Z span (e.g., MW versus SMC or even lower Z).}


\subsection{Ionizing fluxes}
\label{ss_flux}

The ionizing radiation from massive stars has a tremendous impact on their surroundings. As such, it is a key concept that connects stellar and extragalactic astrophysics subjects. A proper interpretation of stellar populations in local and distant starburst galaxies, for example, relies heavily on the their massive star content, evolution, and corresponding ionizing photons output. In a cosmological context, the reionization of the Universe ($z \sim 10-12$) is also linked to the amount of ionizing photons produced by the first hot massive stars, a fraction of which was able to escape their host galaxies \citep[see the review by][]{eldridge22}. 

Robust predictions require computationally expensive, heavily blanketed NLTE expanding atmosphere models. Both the amount of metals and wind presence affect the shape of the spectral energy distribution, changing the number of ionizing photons. Our LMC and SMC grid models were tailored with these considerations.

We have computed the number of ionizing photons Q for the ionization thresholds of \ion{H}{i}, \ion{He}{i}, and \ion{He}{ii}, for all models. Q is defined as follows:

\[ Q = \int _{\nu_o} ^{+\infty} \frac{L_{\nu}}{h\nu} d\nu, \]

 where $\nu_o$ is the frequency threshold for ionization (e.g., of \ion{H}{i}), $L_{\nu}$ is the stellar luminosity at frequency $\nu$, and $h$ the Planck constant.


In Fig. \ref{fig:logq}, we present log q (= log Q$/4\pi R^2$) against the effective temperature, which is the main parameter dictating the peak of the flux distribution (see Fig. \ref{fig:panview}). We chose to present q's (per unit surface area) instead of Q's to ensure a meaningful  comparison between models. Note that for a few temperatures we have just LMC or SMC models (e.g., at 28000 K). 

The number of ionizing photons naturally increases with temperature, as expected. The behavior of ionizing fluxes as a function of metallicity have been discussed by \citet{mp21} and we refer the reader to their Section 4.1 for details. Briefly we see that q(HI) does not change with metallicity, while q(HeI) reaches slightly larger values at lower Z. 

The behavior of q(\ion{He}{ii}) is more complex. \citet{schmutzhamann86}, \citet{hillier87}, \citet{gabler89} and \citet{sdk97} provide an explanation for the sensitivity of \ion{He}{ii} ionizing fluxes to stellar parameters. \ion{He}{ii} opacity depends on the ground state population that is controlled by temperature and wind density. For high temperatures and low wind density, \ion{He}{ii} is fully ionized and the ground state population is low: there is little absorption and the \ion{He}{ii} flux is high. If wind density increases, the line formation depth is pushed outward in regions where the velocity field is non negligible. The \ion{He}{ii} resonance line is shifted so it can absorb more photons, depopulating even more the ground level: the \ion{He}{ii} flux increases accordingly. When the wind becomes strong enough, recombinations dominate over the latter effect and the ground state becomes over-populated: the \ion{He}{ii} flux is severely reduced. This general effect has been described by \citet{sdk97} and is illustrated in Fig. 13 of \citet{mp21}. It explains the bimodal distribution of q(\ion{He}{ii}) at LMC metallicity: depending on the mass-loss rate, the \ion{He}{ii} ionizing flux is either high or low. This can be seen in Fig. \ref{fig:logq}, at \teff\ $\sim$ 40kK,  for the LMC points (rightmost panel). At SMC metallicity, winds remain weak enough in our grid so that the bimodal distribution is barely seen. Obviously, at lower \teff\ ionization is not high enough and the \ion{He}{ii} opacity is always large. Given this behavior, we stress that the predicted \ion{He}{ii} depend on the assumptions regarding mass loss and clumping. The presence of X-ray emission caused by shocks in the winds may also affect the \ion{He}{ii} ionizing fluxes. Hence the absolute values of q(\ion{He}{ii}) we provide should be regarded as only indicative.

\section{Summary and conclusions}
\label{s_conclusions}

We have presented two grids of synthetic spectra computed with the NLTE atmosphere code CMFGEN. These grids cover 
the Main Sequence (and slightly beyond) of massive stars 
at metallicities of $1/2~Z_\odot$ (LMC) and $1/5~Z_\odot$ (SMC).
They serve as an alternative to existing ones in the literature,
built with different codes \citep[e.g.,][]{lh03, lh07, todt15, bpass}.

The models are comprehensive in terms of atomic data, with 
an appropriate amount of metal line blanketing for 
the analysis of OB stars (see Sect. \ref{ss_atom}). A total of
606 spectra were calculated, from 30~\AA\ to 3~$\mu$m. We
provide flux-calibrated as well as normalized spectra in the UV,
optical, and infrared ranges. They are publicly available at the
POLLUX database. 

In addition to spectroscopic data, we have computed photometry
in various filters, including Johnson UBV, Cousins RI, Bessel
JHK, JWST wide, Gaia, and LSST ugrizy filters.
We provide bolometric corrections for all bands and their
calibration as a function of effective temperature. Finally,
we computed \ion{H}{i}, \ion{He}{i}, and \ion{He}{ii} ionizing
fluxes for all models.

The results and data described here are relevant for the quantitative analysis of the UV and optical spectra provided by the \ullyses\ and \xshootu\ observational programs   and other future initiatives 
(e.g., infrared). They can also be used in population synthesis models, to study the nebular properties of \ion{H}{ii} regions, to characterize new populations of massive OB stars uncovered by photometric surveys, 
and on hydrodynamic calculations. Our work also serves as a starting point for developing CMFGEN-based grids at even lower metallicities.


\begin{acknowledgements}
This research has made use of the Spanish Virtual Observatory (\url{https://svo.cab.inta-csic.es}) project funded by MCIN/AEI/10.13039/501100011033/ through grant PID2020-112949GB-I00.

\end{acknowledgements}

\bibliographystyle{aa}
\bibliography{biblio}

\appendix
\section{Appendix}
\label{sec:appendix}

\subsection{UV spectra}
\label{uvspectra}

\begin{sidewaysfigure*}
  \centering
  \includegraphics[width=\textwidth]{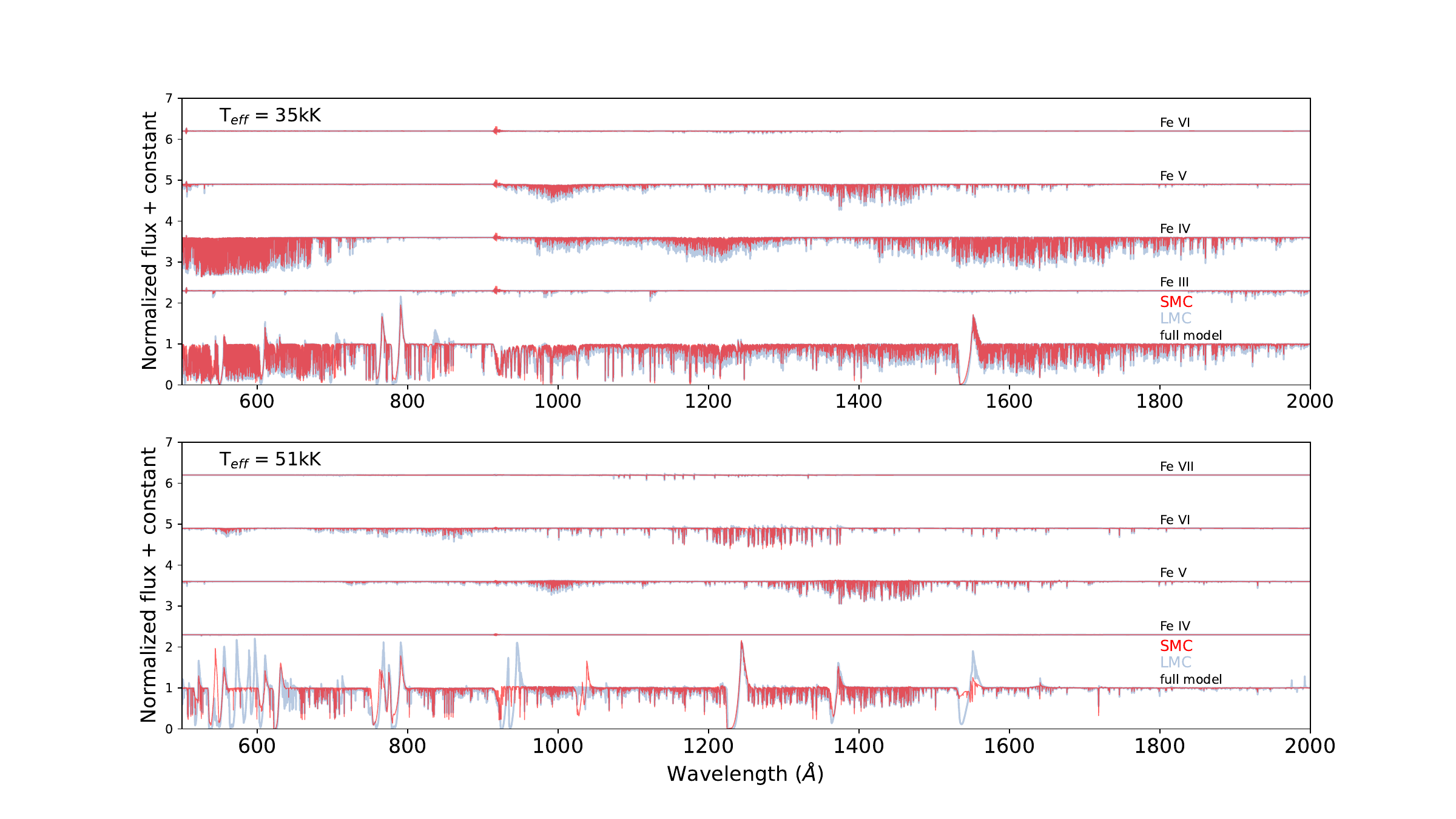}  
  \caption{Ultraviolet spectra: iron forest and metallicity effect. Two models for each temperature are presented: one with 0.5 Z$_\odot$ (LMC; gray) and the other with 0.2 Z$_\odot$ (SMC; red). The surface gravity is log $g$ = 4.0 and the luminosity is log $L/L_\odot = 4.8$ for the \teff\ = 35 kK models. The surface gravity is log $g$ = 4.0 and the luminosity is log $L/L_\odot = 5.9$ (SMC) and log $L/L_\odot = 6.1$ (LMC) for the \teff\ = 51 kK models. Wind parameters are adopted using the Vink formula \citep[][]{vink01}. 
  The full spectra are displayed at the bottom of each panel, while individual contributions of iron ions are shown above. Note the variations in the prevalence of iron ions and the significant reduction in absorption between the SMC and LMC models. }
  \label{fig:uvspectra}
\end{sidewaysfigure*}

Figure \ref{fig:uvspectra} displays examples of spectra in the UV/EUV range, to highlight the role and the richness of the iron forest, and the effects of differing metallicities.


For each metallicity, we selected two effective temperatures, namely \teff\ = 35000 K and \teff\ = 51000 K.
For visualization purposes, the spectra were normalized and convolved with an arbitrary  rotational velocity of 30 km s$^{-1}$. The wavelength interval is 500 -- 2000\AA. Besides the full spectra at the bottom of the two panels, we provide the spectra of individual iron ions, revealing their respective contributions.


At \teff\ = 35000 K, \ion{Fe}{iv-v} dominate. At \teff\ = 51000 K, the wind ionization is higher and \ion{Fe}{v-vi} dominate.
Figure \ref{fig:uvspectra} clearly shows the significant reduction in absorption strengths of iron lines at lower metallicity (SMC).
These plots suggest that a careful analysis of the variation of \ion{Fe}{} ions and lines with temperature might provide a useful tool to constrain \teff\ \citep[see][for a first attempt]{heap06}, i.e.,  independently from the usual helium and/or silicon optical lines in OB stars \citep[see e.g.,][]{almeida19, matheus2023, Crowther06}. 


The P-Cygni profiles of \mbox{CNO} lines are obviously affected by the global model metallicity (the metal abundances are scaled-solar).
They present important variations 
(see for instance, 
\ion{C}{iv} $\lambda$1549 at \teff\ = 51 kK). However, we recall that the models have been calculated with wind parameters obtained using the Vink formula \citep[][]{vink01}, which implicitly scales \mdot\ and \vinf\ to the adopted metallicity. On the other hand, the Ly$\alpha$ wind line remains unchanged, being from hydrogen and saturated (bottom panel).

\subsection{Optical spectra}

\begin{figure*}[htbp]
  \centering
  \includegraphics[width=\linewidth]{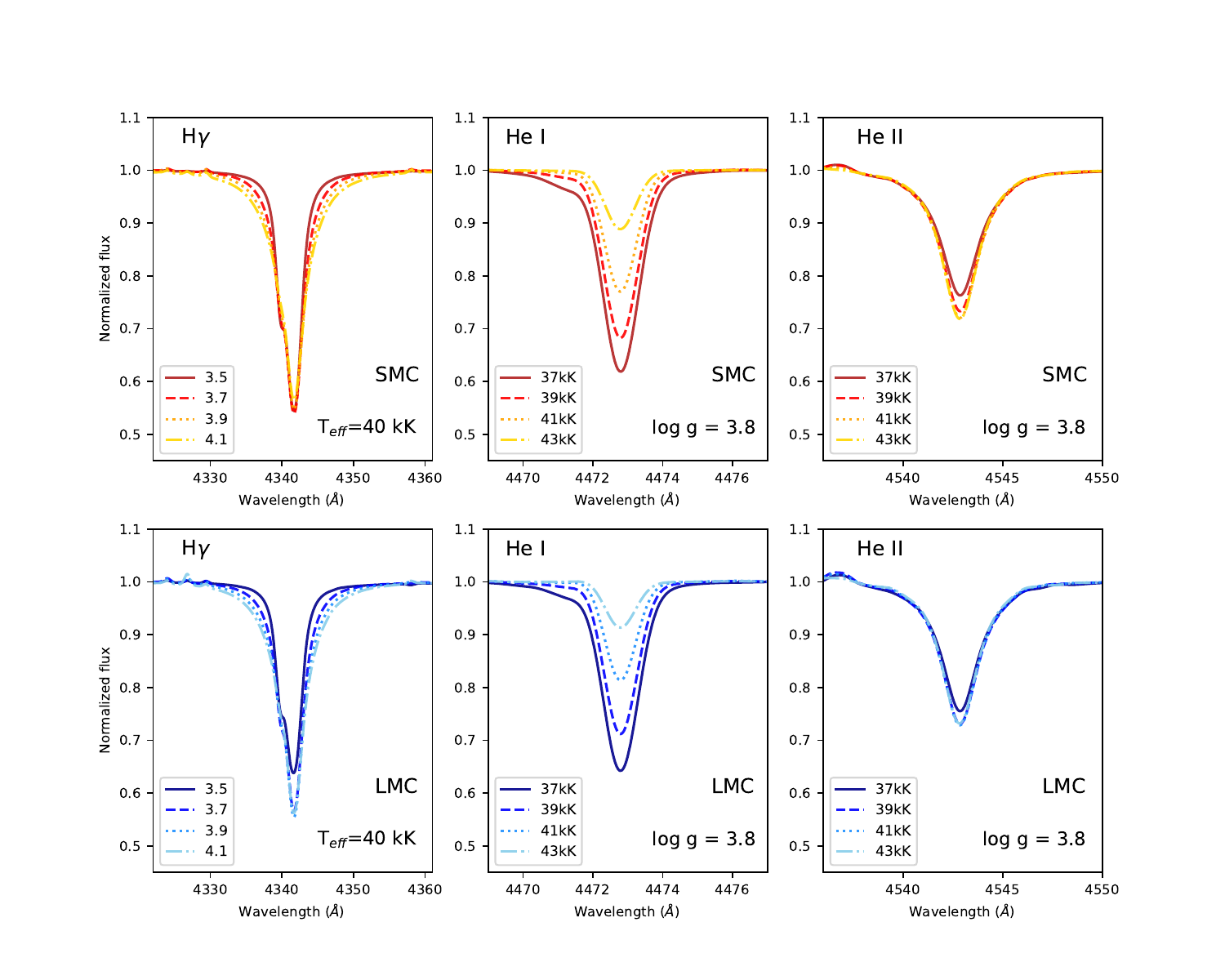} 
  \caption{Classical effects arising from variations in log $g$ and \teff\ on optical spectra, as depicted by our grid models. Top panel: SMC models. Lower panel: LMC models. The chosen helium lines are: \ion{He}{i} $\lambda 4471$ and \ion{He}{ii} $\lambda 4542$. They are often used with other \ion{He}{i-ii} lines to infer \teff\ in O stars, usually with a 1-2 kK uncertainty. The surface gravity determination from H$\gamma$ (and other Balmer lines) has a typical uncertainty of 0.1-0.2 dex. The effective temperature is fixed in the leftmost panels and the \logg\ is varied (see caption). The other panels have \logg\ fixed and the effective temperature is varied (see caption).} 
  \label{fig:logteff-lines}
\end{figure*}

The optical spectra of our grids can be used to infer the surface gravity and effective temperature of an observed star. We present in Fig. \ref{fig:logteff-lines} examples of models with a fixed surface gravity and various \teff\ values and vice-versa. The spectra were convolved with a resolving power $R = 5000$. We selected the lines H$\gamma$, \ion{He}{i} $\lambda 4471$, and \ion{He}{ii} $\lambda 4542$. They have been canonically used for years as diagnostic lines in O stars\footnote{For B stars, \ion{Si}{ii-iv}, \ion{He}{i}, and \ion{Mg}{ii} lines can be used to infer the effective temperature, while Balmer line wings allows the determination of \logg\ \citep[see e.g.,][]{Crowther06, Searle08, matheus2023}}. However, ideally, various helium and hydrogen lines should be utilized simultaneously. 

 The leftmost panels illustrates the changes on the wings of H$\gamma$ when the surface gravity is varied. The effective temperature is fixed at 40 kK. The \logg\, goes from 3.5 to 4.1 in steps of 0.2 dex. When compared with a high signal-to-noise observed spectrum, the changes observed are usually enough for the choice of a best model. An accuracy of 0.1-0.2 dex is commonly obtained in the modeling of high-resolution spectra \citep[see e.g.,][]{bouret12, bouret13}. 

The middle and rightmost panels show the effect of varying the effective temperature. The surface gravity is fixed at \logg\ $= 3.8$. The \ion{He}{i} line changes drastically, while the \ion{He}{ii} changes are more modest. In practice, \teff\, can be inferred with 1-2 kK accuracy \citep[see e.g.,][]{bouret12,bouret13}.

\begin{figure*}[htbp]
  \centering
  \includegraphics[width=0.9\linewidth]{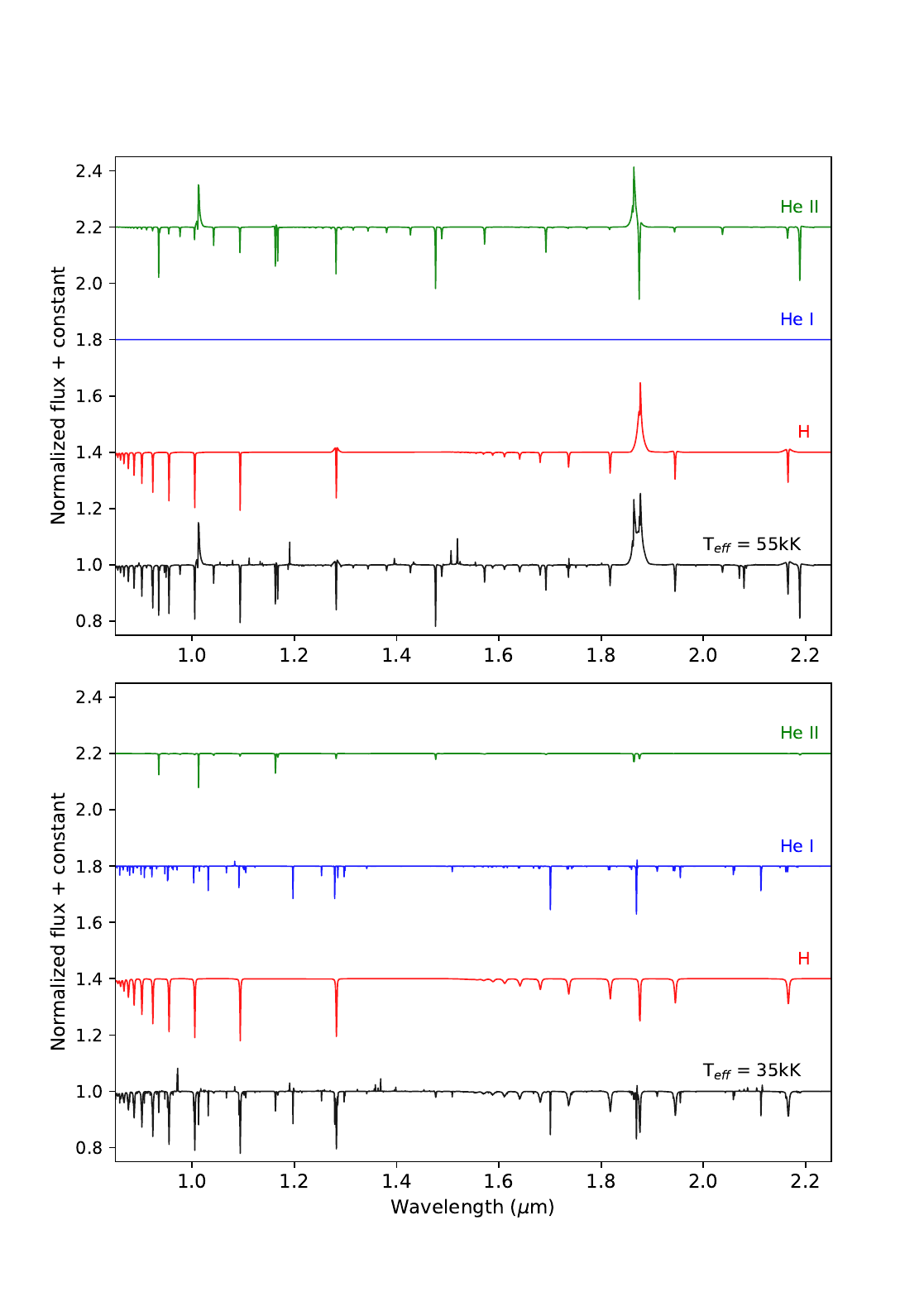} 
  \caption{Examples of infrared spectra at 0.85--2.25 $\mu$m (LMC metallicity). Top panel: model with \teff\ = 55 kK and log $g$ =  4.0. Bottom panel: model with \teff\ = 35 kK and log $g$ =  4.0. The spectra were convolved with R = 2700, appropriate, for example, for \jwst\ NIRSpec data. In each panel, the full spectrum is shown at the bottom (black) and individual \ion{H}{i}, \ion{He}{i}, and \ion{He}{ii} contributions are denoted above. } 
  \label{fig:ir-lines}
\end{figure*}

\subsection{Infrared spectra}

We present selected infrared spectra in Fig. \ref{fig:ir-lines}. Here, we chose only LMC models. 
The wavelength interval is 0.85--2.25 $\mu$m. We convolved all spectra with R=2700, for visualization purposes. This resolving power is, for instance, compatible with the \jwst\ NIRSpec instrument in this spectral range (disperser-filter G140H/F100LP and G235H/F170LP). Two effective temperatures are shown, namely, 35kK and 55kK. The surface gravity is \logg\ $= 4.0$. Individual contributions from hydrogen and helium are displayed above the full spectrum in each panel. Most hydrogen lines correspond to the Paschen ($j \rightarrow$ 3) and Brackett ($j \rightarrow$ 4) series. Note the shift in the \ion{He}{i}/\ion{He}{ii} ratio in the two temperatures. High temperature models do not possess \ion{He}{i} transitions in the wavelength interval shown.


\end{document}